\documentclass[12pt]{article}

\usepackage{color}
\usepackage{epsfig,amssymb,amsfonts,amsmath,graphicx,dsfont,cite,xfrac}
\usepackage{authblk}
\usepackage{subcaption}
\definecolor{mygray}{gray}{0.5}

\usepackage{cite}
\usepackage[colorlinks=true,linkcolor=blue,citecolor=red]{hyperref}
\usepackage{soul,cancel}

\newcommand{\be}{\begin{equation}}
\newcommand{\ee}{\end{equation}}
\newcommand{\bea}{\begin{eqnarray}}
\newcommand{\eea}{\end{eqnarray}}

\title{Optimizing entanglement in two-qubit systems}

\author[${1}$]{Salvio Luna-Hern\'andez}
\author[${2}$]{Claudia Quintana}
\author[${1,{}^*}$]{Oscar Rosas-Ortiz}

\affil[${1}$]{\footnotesize Physics Department, Cinvestav, AP 14-740, 07000
M\'exico City, Mexico}

\affil[${2}$]{\footnotesize School of Engineering and Sciences, 
Tecnol\'ogico de Monterrey, Atizap\'an 52926, Mexico}

\affil[${^*}$]{\footnotesize Correspondence: oscar.rosas@cinvestav.mx}

\date{}
\begin{document}

\maketitle

\begin{abstract}
We investigate entanglement in two-qubit systems using a geometric representation based on the minimum of essential parameters. The latter is achieved by requiring subsystems with the same entropy, regardless of whether the state of the entire system is pure or mixed. The geometric framework is provided by a convex set $\mathcal S$ that forms a right-triangle, whose points are linked to just two of the coherences of the system under study. As a result, we find that optimized states of two qubits are X--shaped and host pairs of identical populations while reducing the number of coherences involved. A geometric $L$-measure of entanglement is introduced as the distance between the points in $\mathcal S$ that represent entangled states and the closest point that defines separable states. It is shown that $L$ reproduces the results of the Hill-Wootters concurrence $C$, so that $C$ can be interpreted as a distance-like entanglement measure. However, unlike $C$, the measure $L$ also distinguishes the rank of states. The universality of the two-qubit X--states ensures the utility of our geometric model for studying entanglement of two-qubit states in any configuration. To show the applicability of our approach far beyond time-independent cases, we construct a time-dependent two-qubit state, traced out over the complementary components of a pure tetra-partite system, and find that its one-qubit states share the same entropy. The entanglement measure results bounded from above by the envelope of the minima of such entropy.
\end{abstract}

%%%%%%%%%%%%%%%%%%%%%%%%%%%%%%%%%%%%%%%%

%---------------------------------------> Section
\section{Main}
\label{intro}

Quantum correlations characterize the features of quantum multipartite systems that cannot be described by classical--like theories. Closely related to the concept of quantum coherence, no quantum correlation is viable without interference with respect to distinguishable alternatives of quantum states. Nowadays nonclassical correlations are invaluable resources for quantum information and quantum computation \cite{Nie00}, but they emerged on the scene by raising one of the most important problems in the foundations of quantum theory \cite{Ein35,Sch35} and, surprisingly, by providing the solution of the same \cite{Bel64,Asp81}. Schr\"odinger introduced the expression {\em entanglement} for this first manifestation of quantum correlation \cite{Sch35}. The concept describes what occurs with our knowledge of two systems that are separated after they were interacting for a while, and from which we had maximal knowledge before the interaction \cite{Ros19}. In this sense, entanglement implies that the state of a composite system cannot be separated into the states of its parts. However, entanglement does not account for all quantum correlations, and even separable states may exhibit correlations that are not completely classical. 

The entanglement in a system can be characterized and quantified from different points of view \cite{Ved97,Vid99,Hor01,Bla08,Hor09,Guh09,Ple14,Kry21,Zha22,Lun23a,Lun23,Lun24}. However, the separability criterion introduced by Peres \cite{Per96} and the Horodecki family \cite{Hor96} provides a powerful tool  for its detection. In this work, the Peres-Horodecki (also known as positive partial transpose --PPT--) criterion is applied to construct a convex representation of two-qubit states that permits the study of entanglement in purely geometric form. It is well known that this criterion is necessary and sufficient to detect entanglement in $2 \otimes 2$ and $2 \otimes 3$ quantum systems \cite{Hor96}, so we will use it to identify entanglement in two-qubit mixed states. There are several incentives to consider these states, mainly because generating pure entangled states is not an easy task, so entangled states could be prepared as statistical mixtures for purely experimental reasons, but in general they can be the result of a partial trace of the quantum state of a larger system. 

In general, mixed states characterize the interaction between the system and its environment, and the study of their entanglement properties is not only more complicated than that of pure states, but lesser understood. Although the first entangled system found in physics was in a pure state \cite{Ein35}, it is in one of the Bell states \cite{Bel64}, shortly after mixed states were identified that also disrupt the foundations of quantum theory \cite{Wer89}. These states, named after Werner, can be modeled using hidden variables without violating the Bell inequalities, and find applications in a diversity of quantum entanglement subjects \cite{Lee00,Hir00,Pit00,Lyo12,Dia18,Gha21,Jir21,Abo23,Zha24,Jia24}. Our results include a wide class of Werner states as a particular case.

We investigate the entanglement properties of two-qubit systems by using the minimum of essential parameters. The idea is to construct density operators whose reduced one-qubit states share the same entropy, regardless of whether the state of the entire system is pure or mixed. In this way, we identify X--shaped states \cite{Mun01,Yu04,Yon06,Wan06,Yu07,Que12,Men14}, denoted $\rho_X$, that have the capacity to host pairs of identical populations while reducing the number of coherences involved. The X--states played a significant role to show that nonlocal disentanglement times are shorter than local decoherence times for spontaneous emission of two initially entangled qubits \cite{Yu04} (see also \cite{Yon06}). One of the most relevant things to note is that concurrence takes a very simple form for X--states \cite{Wan06,Yu07,Que12}, which justifies the extensive application of these states to study the dynamics of entanglement between two qubits \cite{Yu04,Yon06,Wan06,Yu07,Que12,Men14,Qas08,Ali09,Enr14,Qui15,Qui16}. Demanding positive semidefinite, self-adjoint and normalized density operators $\rho_X$, and imposing the Peres-Horodecki criterion, we arrive at the system of inequalities that related matrix elements must satisfy to determine both, the eligibility of $\rho_X$  as a quantum state, and the possibility of finding entanglement \cite{Que12}. We show that the solution of such a system defines a convex set $\mathcal S$ that can be divided in subsets whose points  are linked to separable states and regions where entanglement may be found. 

Based on the convex representation  provided by $\mathcal S$, we introduce a geometric $L$-measure of entanglement, a distance-like entanglement measure \cite{Aul10}, that reproduces the results of the Hill-Wootters generalized concurrence $C$ \cite{Hil97,Woo98} when applied to  states $\rho_X$. Thus, our results give $C$ a geometric meaning that is not obvious without the convex structure $\mathcal S$. However, unlike $C$, the measure $L$ distinguishes the rank of states with maximum entanglement. The latter is relevant to determine not only the amount of quantum correlations but rather the quantumness of the state \cite{Ges12}, so it shows how much useful a state is for a specific quantum processing task \cite{Wan13,Man15}. Another advantage of the geometric $L$-measure is that, with respect to other entanglement measures, it allows comparatively easy calculation.

The universality of two-qubit X--states was first explored numerically \cite{Hed13}. There, it is presented `strong numerical evidence showing that any general state can indeed be transformed to an X state of the same entanglement'. It has later been shown that X--states are in fact universal \cite{Men14}. That is, given a two-qubit density operator $\rho$, a unitary transformation can be found that produces an X--shaped state $\rho_X$ with equal (or higher) entanglement than $\rho$. The spectrum and purity of $\rho$ are inherited unchanged to $\rho_X$. In this way, the geometric formulation we develop throughout this work for two-qubit X--states is useful for any configuration of the two-qubit density operators.

We also study what happens to entanglement when the system depends explicitly on time. Modeling an entangled diatomic system whose one-qubit parts are placed in independent, isolated and identical electromagnetic cavities, each of which contains $n$ photons, we construct a tetra-partite pure state that continues to evolve unitarily as a pure state. However, the reduced diatomic state does not evolve in a reversible way (systems like this may show signatures of chaos \cite{Nag01}), which makes it an excellent subject of study to understand the behavior of entanglement in systems described by time-dependent statistical mixtures. Regardless of the dynamical law obeyed by the time-dependent diatomic state, the corresponding one-qubit states share the same von Neumann entropy $S$, which oscillates as time passes. Our results verify that finding reduced states of maximum mixing does not provide information about the entanglement (if any) of the entire system when it is in a mixed state. However, the envelope formed by the minima of $S$ defines an upper bound for the entanglement measure. To the best of our knowledge, this remarkable result has gone unnoticed in the literature on the matter.

%---------------------------------------> Section
\section{Two-qubit states}
\label{2qubit}

The density operator of any two-qubit system can be written as a $4 \times 4$ complex matrix over the field $\mathbb C$,
\be
\rho = \left( r_{kj} \right), \quad r_{jk} = r^*_{kj} \in \mathbb C, \quad k,j \in \{1,2,3,4\}, \quad \sum_{k=1}^4 r_{kk} =1,
\label{state}
\ee
where $z^*$ stands for the complex-conjugate of $z \in \mathbb C$. The populations (diagonal elements) and coherences (off-diagonal elements) are such that $\rho$ is self-adjoint ($\rho= \rho^{\dagger}$), normalized ($\operatorname{Tr} \rho =1$) and positive semidefinite ($\rho \geq 0$). The condition $\operatorname{Tr} \rho^2 = 1$ is necessary and sufficient for a density operator $\rho$ to represent a pure state. In turn, $\rho$ refers to mixed states if $\operatorname{Tr} \rho^2 < 1$. Thus, in general, matrix (\ref{state}) is defined by 15 real parameters.

Hereafter,  to represent states in the Hilbert space $\mathcal{H} = \mathbb C^4$, we use the computational basis $\vert 00 \rangle \equiv \vert e_1 \rangle$, $\vert 01 \rangle \equiv \vert e_2 \rangle$, $\vert 10 \rangle \equiv \vert e_3 \rangle$, $\vert 11 \rangle \equiv \vert e_4 \rangle$, with $\vert e_1 \rangle = (1,0,0,0)^T$, and so on.

Assuming that the matrix-representation of $\rho$ is known in advance, there is a simple way to identify whether or not it represents a pure state, which consists of the applicability of the following proposition (the proof is provided in Appendix~\ref{ApA}).

{\bf Proposition~Q.}  If the matrix-elements of the two-qubit density operator $\rho = \left( r_{kj} \right)$ can be factorized in the form $r_{kj} = \alpha_k \alpha^*_j$, with $\alpha_k, \alpha_j \in \mathbb C$, and $k,j \in \{1,2,3,4\}$, then $\rho$ represents the pure state $\vert \psi \rangle = \sum_{k=1}^4 \alpha_k \vert e_k \rangle$.

We are interested in density operators whose matrix-elements do not necessarily satisfy Proposition~Q, which opens the possibility of investigating entanglement in two-qubit mixed states. Among the diverse ways to prepare such states, we pay attention to the representation generated by the computational basis.

To outline the profile of the states of our interest, let us consider the reduced density operators $\rho_{1,2} = \operatorname{Tr}_{2,1} \rho$. After diagonalizing, we obtain $\rho_{k,\operatorname{diag}} = \operatorname{diag} \left( \lambda_{k}, 1 - \lambda_{k} \right)$, where $\lambda_{k} = (1- \sqrt{1 - 4 \det \rho_{k}})/2$ is the smallest eigenvalue of $\rho_{k}$. The probability vectors $\vec \lambda_k =  \left( \lambda_{k}, 1 - \lambda_{k} \right)$ do not match in general since $\det \rho_{2} = \det \rho_{1} + \Delta$, with
\begin{equation}
\Delta = (r_{11} - r_{44})(r_{22} - r_{33}) + \vert r_{13} + r_{24} \vert^{2} - \vert r_{12} + r_{34} \vert^{2}.
\label{delta}
\end{equation}
However, if $\Delta=0$ then $\vec\lambda_1 = \vec\lambda_2 = \vec\lambda = (\lambda, 1 - \lambda)$ and $\rho_{1,\operatorname{diag}}=\rho_{2,\operatorname{diag}}$. In such a case the von Neumann entropy gives the same result for both reduced matrices $S(\rho_1) = S (\rho_2)$.

Throughout this work the search for subsystems with equal entropy serves as a way to construct two-qubit mixed states that require a reduced number of parameters to characterize their possible entanglement. As long as we have these states, we will establish a mechanism to identify (and measure) entanglement.

With respect to the degree of mixing, the matrix rank of $\rho$ (the number of its nonzero eigenvalues) is quite useful since it identifies the minimum number of pure states needed to prepare the corresponding statistical mixture. In fact, this notion is directly related to the purity of $\rho$, $\tfrac14 \leq \eta = \operatorname{Tr} \rho^2 \leq 1$, since a pure state has rank 1 (and purity 1). So that purity less than 1 implies rank greater than 1. In this way, the maximally mixed state $\rho_{\star}= \tfrac14 \mathbb I$ has rank 4 (and purity 1/4).

A very simple form to satisfy $\Delta =0$ is by imposing the constraint
\be
r_{13}= r_{24} = r_{12} = r_{34} =0,
\label{rule1}
\ee
which yields diagonal reduced matrices, together with at least one of the following conditions
\be
r_{11}= r_{44}, \quad r_{22}= r_{33}.
\label{rule2}
\ee
Note that (\ref{rule1}) removes 8 parameters from $\rho$ while (\ref{rule2}) cancels at least one more parameter. In this way, as long as $\Delta=0$ is true, a maximum of 6 real parameters will suffice to define the quantum states $\rho$ that we study below.

%---------------------------------------> Subsection
\subsection{Optimized states of two qubits are X--shaped}
\label{Xstates}

To construct states that satisfy the $\Delta = 0$ requirement, let us first consider that only condition (\ref{rule1}) is satisfied. In this case the density operators (\ref{state}) require only 7 real parameters and are X--shaped,
\begin{equation}
\rho_{X} = \left(
\begin{array}{cccc}
r_{11} & 0 & 0 & r_{14}\\
0 & r_{22} & r_{23} & 0\\
0 & r_{23}^{*} & r_{33} & 0\\
r_{14}^{*} & 0 & 0 & r_{44}
\end{array}
\right).
\label{rhox}
\end{equation}

Since coherence is a basis-dependent concept, states represented by non-diagonal density operators contain a certain amount of coherence with respect to the chosen basis. To be concrete, as density operators (\ref{state}) are written in terms of the computational basis, their nonzero coherences $r_{kj}$, $k\neq j$, determine the possibility of quantum interference with respect to the distinguishable alternatives $\vert e_k \rangle$ and $\vert e_j \rangle$. In this way, for X--states (\ref{rhox}), interference is only allowed between $\vert e_1 \rangle = \vert 00 \rangle$ and $\vert e_4 \rangle = \vert 11 \rangle$, as well as between $\vert e_2 \rangle = \vert 01 \rangle$ and $\vert e_3 \rangle = \vert 10 \rangle$. Any other interference between the elements of the basis is not allowed. In the most general case both interferences occur simultaneously, but entanglement arises when at least one of them is activated. 

To analyze the properties of X--states as generally as possible, it is necessary to ensure that $\rho_X$ is in fact a density operator. By construction, $\rho_X$ is self-adjoint and normalized, so we just need to make sure that it is positive semidefinite. 

After calculating the eigenvalues of $\rho_X$, see explicit expressions in Eq.~(\ref{lambdas}) of Appendix~\ref{ApA}, we find that they are non-negative as long as the following conditions are met \cite{Que12}:
\begin{equation}
\vert r_{14} \vert \leq \sqrt{r_{11} r_{44}}, \qquad \vert r_{23} \vert \leq \sqrt{r_{22} r_{33}}.
\label{possemi}
\end{equation}
That is, $\rho_X$ represents an admissible quantum state, pure or mixed, only if the coherence amplitudes $\vert r_{14} \vert$ and $\vert r_{23} \vert$ are upper bounded by the population of the corresponding states. Note that the phases of $r_{14}$ and $r_{23}$ play no role in characterizing $\rho_X$ as a positive semidefinite matrix. Nevertheless, they characterize quantum interference with respect to the corresponding distinguishable alternatives.

%---------------------------------------> Section
\subsubsection{Subsystems with equal entropy}
\label{equal}

Assuming that not only (\ref{rule1}) but the first identity of Eq.~(\ref{rule2}) is also satisfied, states (\ref{rhox}) will have the same population for states $\vert e_1 \rangle = \vert 00 \rangle$ and $\vert e_4 \rangle = \vert 11 \rangle$. As a consequence, the reduced states will have the same probability vector $\vec \lambda_L = (\lambda_L, 1- \lambda_L)$, with $\lambda_L = (1 -\vert r_{22} -r_{33} \vert )/2$ and  $2r_{11} + r_{22}+r_{33}=1$. 

On the other hand, satisfying (\ref{rule1}) and the second identity of Eq.~(\ref{rule2}) leads to states (\ref{rhox}) characterized by having the same population for states $\vert e_2 \rangle = \vert 01 \rangle$ and $\vert e_3 \rangle = \vert 10 \rangle$. The probability vector is now $\vec \lambda_R = (\lambda_R, 1- \lambda_R)$, with $\lambda_R = (1 -\vert r_{11} -r_{44} \vert)/2$ and $r_{11}+ 2r_{22} + r_{44} =1$. 

Since populations satisfy $0 \leq r_{kk} \leq 1$, we obtain $0 \leq \lambda_{L,R} \leq \frac12$. That is, in general, neither of the two previous cases maximizes entropy. 

Reduced matrices with maximal entropy are achieved if, in addition, we make $r_{22}= r_{33}$ in $\lambda_L$ and $r_{11}= r_{44}$ in $\lambda_R$. The latter corresponds to also satisfying the complementary identity of Eq.~(\ref{rule2}) in each case. That is, only if (\ref{rule1}) and both identities of (\ref{rule2}) are satisfied, one gets $\lambda_L = \lambda_R = 1/2$.

The above reasoning applies to $\rho_X$ matrices regardless of whether they represent pure or mixed states. Therefore, when (\ref{rule1}) and both identities in (\ref{rule2}) hold, the density operators (\ref{rhox}) are optimized and host pairs of identical populations. This further reduces the number of parameters needed to define $\rho_X$ from 7 to just 5.

With respect to the pure states represented by the optimized $\rho_X$ matrices, in Appendix~\ref{ApA} we show that once Proposition~Q is satisfied, when (\ref{rule1}) and both identities of (\ref{rule2}) hold, we recover the Bell--states 
\be
\vert \beta_1 \rangle = \frac{ \vert e_1 \rangle + \vert e_4 \rangle}{\sqrt 2}, \qquad 
\vert \beta_2 \rangle = \frac{\vert e_1 \rangle - \vert e_4 \rangle}{\sqrt 2},
\label{bell1}
\ee
and
\be
\vert \beta_3 \rangle = \frac{\vert e_2 \rangle + \vert e_3 \rangle}{\sqrt 2}, \qquad 
\vert \beta_4 \rangle = \frac{\vert e_2 \rangle - \vert e_3 \rangle}{\sqrt 2}.
\label{bell2}
\ee
So the Bell basis elements (\ref{bell1})-(\ref{bell2}) are optimized states of two qubits.

To give a more general example, consider the convex combinations
\be
\rho_B = \sum_{k=1}^4 b_k \vert \beta_k \rangle \langle \beta_k \vert, \quad 0 \leq b_k \leq 1, \quad \sum_{k=1}^4 b_k =1.
\label{rhob}
\ee
This family of density operators is not only X--shaped; their members accommodate two pairs of equally weighted populations and reduce the number of coherences. Namely, the density operators $\rho_B$ are optimized:
\begin{equation}
\rho_B = \frac12 \left(
\begin{array}{cccc}
b_1 + b_2 & 0 & 0 & b_1-b_2\\
0 & b_3+b_4 & b_3 -b_4& 0\\
0 & b_3-b_4 & b_3 + b_4 & 0\\
b_1 -b_2 & 0 & 0 & b_1+b_2
\end{array}
\right).
\label{rhobm}
\end{equation}
The one-qubit states that make up any state $\rho_B$ are maximally mixed $S(\rho_{B,1}) = S(\rho_{B,2}) = 1$. This set of optimized two-qubit states includes two special configurations, the first reduces $\rho_B$ to a single Bell--state ($b_k=1$ for a given $k$), $\rho_B = \vert \beta_k \rangle \langle \beta_k \vert$, and the second yields an equally weighted statistical mixture of Bell--states ($b_k = \frac14$ for any $k$), $\rho_B= \rho_{\star}$. 

For states $\rho_B$, the coherences $r_{14} =r_{14}^* =b_1 -b_2$ and $r_{23} =r_{23}^* =b_3 -b_4$ are activated whenever $b_1 \neq b_2$ and $b_3 \neq b_4$, respectively. This property allows us to identify other interesting cases of optimized two-qubit states. For example, interference between $\vert e_2 \rangle = \vert 01 \rangle$ and $\vert e_3 \rangle = \vert 10 \rangle$ is not allowed if $r_{23} = 0$, which can be achieved by setting $b_3=b_4 =\kappa$ in (\ref{rhob}) to get
\be
\left. \rho_B \right\vert_{b_3=b_4=\kappa}  = b_1 \vert \beta_1 \rangle \langle \beta_1 \vert + b_2 \vert \beta_2 \rangle \langle \beta_2 \vert + \kappa \left( \vert 01 \rangle \langle 01 \vert + \vert 10 \rangle \langle 10 \vert\right), \quad b_1+b_2+2\kappa=1.
\label{rhobm2}
\ee
To avoid reaching a statistical mixture of the computational basis, we will require $b_1 \neq b_2$. The matrix representation is easily obtained from (\ref{rhobm}) and reads as follows
\begin{equation}
\left. \rho_B \right\vert_{b_3=b_4=\kappa}= \left(
\begin{array}{cccc}
\frac{b_1 + b_2}{2} & 0 & 0 & \frac{b_1-b_2}{2}\\
0 & \kappa &0 & 0\\
0 & 0 & \kappa & 0\\
\frac{b_1 -b_2}{2} & 0 & 0 & \frac{b_1+b_2}{2}
\end{array}
\right), \quad b_1 \neq b_2, \quad b_1+b_2 +2\kappa=1.
\label{rhobm3}
\end{equation}
As we can see, what distinguishes this state from a statistical mixture of the basis elements is the coherence value $r_{14} = (b_1-b_2)/2$. The greater the interference between $\vert e_1 \rangle = \vert 00 \rangle$ and $\vert e_4 \rangle = \vert 11 \rangle$, the greater the difference.

An important case of the previous model is achieved by setting $b_k = (1+3q)/4$, and $b_j = (1-q)/4$ for $k = \operatorname{fixed}$, and $j \neq k$. This leads to the Werner--states \cite{Wer89}:
\be
\rho_{W_k} = q \vert \beta_k \rangle\langle \beta_k \vert + \left( \tfrac{1 - q}{4} \right) \mathbb I, \quad -\tfrac13 \leq q \leq 1, \quad k=1,2,3,4.
\label{werner}
\ee
Explicitly,
\bea
\rho_{W_j} = \left(
\begin{array}{cccc}
\frac{1+q}{4} & 0 & 0 & \frac{(-)^{j+1} q}{2}\\
0 & \frac{1-q}{4} & 0 & 0\\
0 & 0 & \frac{1-q}{4} & 0\\
\frac{ (-)^{j+1} q}{2} & 0 & 0 & \frac{1+q}{4}
\end{array}
\right), \quad
\rho_{W_{\ell}} = \left(
\begin{array}{cccc}
\frac{1-q}{4} & 0 & 0 & 0\\
0 & \frac{1+q}{4} &  \frac{(-)^{\ell-1} q}{2} & 0\\[1ex]
0 & \frac{(-)^{\ell-1} q}{2} & \frac{1+q}{4} & 0\\[1ex]
0 & 0 & 0 & \frac{1-q}{4}
\end{array}
\right),
\label{rhow}
\eea
where $j=1,2$ and $\ell =3,4$. That is, the Werner--states (\ref{rhow}) are not only X--shaped but they are also optimized in the sense discussed above. 

Comparing (\ref{rhow}) with (\ref{rhobm3}), we see that $\left. \rho_B \right\vert_{b_3=b_4=\kappa}$ has the structure of the Werner--states $\rho_{W_j}$.

The Werner--states $\rho_{W_k}$ can be generalized to the configuration
\be
\rho_S = \sum_{k=1}^4 q_k \vert \beta_k \rangle \langle \beta_k \vert + \left( \tfrac{1 - s}{4} \right) \mathbb I, \quad \sum_{k=1}^4 q_k = s, \quad \tfrac{s-1}{4} \leq q_k \leq \tfrac{s+3}{4},
\label{rhos}
\ee
where 
$ s \leq 1+q_{\min}$ and $q_{\min} = \min \{ q_1, q_2, q_3, q_4\}$. The explicit form of these optimized density operators is as follows
\begin{equation}
\rho_S = \left(
\begin{array}{cccc}
\varrho_{11} & 0 & 0 & \frac{q_1-q_2}{2}\\
0 & \varrho_{22} & \frac{q_3 -q_4}{2} & 0\\
0 & \frac{q_3-q_4}{2} & \varrho_{22} & 0\\
\frac{q_1 -q_2}{2} & 0 & 0 & \varrho_{11}
\end{array}
\right),
\label{rhosm}
\end{equation}
where
\[
\varrho_{11}= \tfrac14 \left( 1 + q_1 +q_2 - q_3 -q_4 \right), \qquad 
\varrho_{22}= \tfrac14 \left( 1 - q_1 -q_2 + q_3 +q_4 \right).
\]

%---------------------------------------> Section
\subsection{Entanglement conditions}
\label{conditions}

The positive partial transposition of the density operator (\ref{state}) obeys the following transformation rules
\[
\begin{array}{c}
\rho^{T_1}: \quad r_{13} \rightarrow r_{13}^*, \, r_{14} \rightarrow r_{23}^*, \, r_{24} \rightarrow r_{24}^*.\\[2ex]
\rho^{T_2}: \quad r_{12} \rightarrow r_{12}^*, \, r_{14} \rightarrow r_{23}, \, r_{34} \rightarrow r_{34}^*.
\end{array}
\]
Without loss of generality, we will use the transposition with respect to the second qubit.

The transposition of the X--state (\ref{rhox}) corresponds to make $r_{14} \rightarrow r_{23}$. Eigenvalues (\ref{lambdas}) and eigenvectors (\ref{evec}) must be transformed accordingly to solve the eigenvalue equation $\rho_X^{T_2} \vert \widetilde \epsilon \, \rangle = \widetilde \Lambda \vert \widetilde \epsilon \, \rangle$. Then, it is found that $\rho_X^{T_2}$ is positive semidefinite whenever $\vert r_{23} \vert \leq \sqrt{r_{11} r_{44}}$ and $\vert r_{14} \vert \leq \sqrt{r_{22} r_{33}}$. Applying the Peres--Horodecki criterion \cite{Per96,Hor96}, we find that $\rho_X$ is entangled as long as one of the following inequalities hold \cite{Que12}:
\begin{equation}
 \vert r_{23} \vert > \sqrt{r_{11} r_{44}} , \qquad \vert r_{14} \vert > \sqrt{r_{22} r_{33}}.
\label{ppt}
\end{equation}
This result imposes a lower limit on the coherence amplitudes that is unexpectedly determined by the populations of the complementary states. That is, to find entanglement, $\vert r_{14} \vert$ must be lower bounded by the population of states $\vert e_2 \rangle = \vert 01 \rangle$ and $\vert e_3 \rangle = \vert 10 \rangle$. Similar conclusions hold for the coherence amplitude $\vert r_{23} \vert$  and populations $r_{11}$ and $r_{44}$.

Thus, once the populations of the X--states are known, the coherences determine both the eligibility of $\rho_X$ as a quantum state and the possibility of finding entanglement.

%---------------------------------------> Section
\section{Convex optimization}
\label{optimize}

To characterize entanglement in the X--states we have found a system of inequalities. Remember, (\ref{possemi}) has been introduced to ensure that $\rho_X$ is positive semidefinite while (\ref{ppt}) obeys the Peres--Horodecki criterion for entanglement. 

\begin{figure}[h!]

\centering 
\includegraphics[width=.25\textwidth]{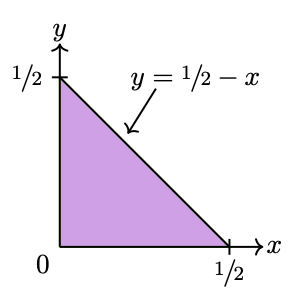} 

\caption{\footnotesize The points of the convex set $\mathcal S \subset \mathbb R^2$, a right-triangle with legs of length 1/2, determine the matrix elements with which the density operator $\rho_X$ becomes positive semidefinite.}

\label{Fig01}
\end{figure}

Using shorter notation we write $x_{0} = \sqrt{r_{11} r_{44}}$, $y_{0} = \sqrt{r_{22} r_{33}}$, $r_{14} = x e^{i \theta}$ and $r_{23} = y e^{i \phi}$, with $0 \leq \theta, \phi < 2 \pi$. In this way, condition (\ref{possemi}) reads 
\begin{equation}
0 \leq x \leq x_{0}, \quad 0 \leq y \leq y_{0}.
\label{rhocond}
\end{equation}
The combined inequalities (\ref{rhocond}) delimit a convex region of the $xy$--plane defined by $0 \leq x + y \leq x_{0} + y_{0}$. The maximum value of the upper bound is optimized as $\left( x_0 +y_0 \right)_{\max}= 1/2$, see details in Appendix \ref{ApA}. Therefore, $\rho_X$ is positive semidefinite for the points contained in the convex set 
\be
\mathcal{S} = \left\{ (x,y) \in \mathbb R^2;  0 \leq x + y \leq \tfrac12 \right\}.
\label{convex}
\ee
Geometrically, $\mathcal S$ is nothing more than the right-triangle shown in Figure~\ref{Fig01}. 

%%%%%%%%%%%%%%%%%%%%%%%%%%%%%%%%%%
%%%%%%%%%%%%%%%%%%%%%%%%%%%%%%%%%%

This is the scenario where at least one of the inequalities (\ref{ppt}) must be satisfied to obtain entanglement. We have two possible configurations, 
\be
0 \leq y \leq y_{0}, \quad  y_0 < x \leq x_{0},
\label{entcrit2}
\ee
or
\be
0 \leq x \leq x_{0}, \quad x_0 < y \leq y_{0}.
\label{entcrit1}
\ee

\begin{figure}[h!]

\centering
\subfloat[][$\rho_{W_{1,2}}$]{\includegraphics[width=0.3\textwidth]{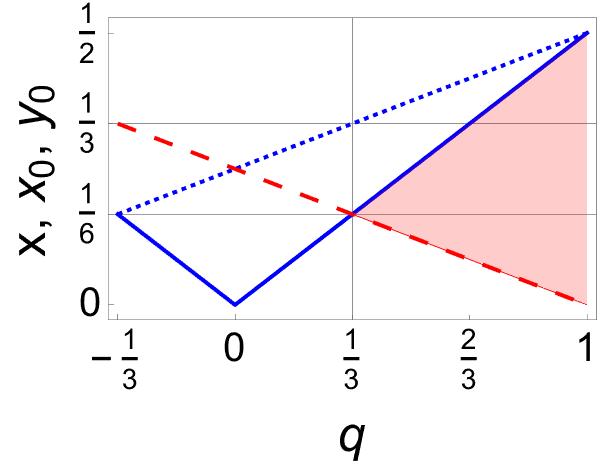}}
\hskip1cm
\subfloat[][$\rho_{W_{3,4}}$]{\includegraphics[width=0.3\textwidth]{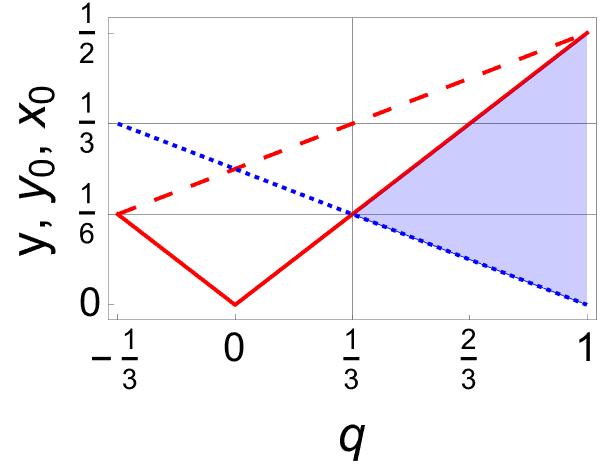}}

\caption{\footnotesize Entanglement conditions for the Werner--states $\rho_{W_k}$, with $x = \vert r_{14} \vert$ (blue-continuous), $y = \vert r_{23} \vert$ (red continuous), $x_0 = \sqrt{r_{11} r_{44}}$ (blue-dotted), and $y_0 = \sqrt{r_{22} r_{33}}$ (red-dashed). Two general cases are distinguished ({\bf a}) If $k=1,2$, then $y=0$. To find entanglement the coherence amplitude $x$ must obey inequalities (\ref{entcrit2}), see shaded-red area ({\bf b}) If $k=3,4$, then $x=0$. Entanglement is found whenever the coherence amplitude $y$ obeys inequalities (\ref{entcrit1}), see shaded-blue area.
}
\label{F02}
\end{figure}

Consider for example the Werner--states (\ref{werner}). If $k=1,2$, we find $y=0$, $x = \frac12 \vert q \vert$, $y_0 = \frac14 (1-q)$ and $x_0 = \frac 14 (1+q)$, see Figure~\ref{F02}(a). To find entanglement, $x$ must obey inequalities (\ref{entcrit2}). If $k=3,4$, the results are obtained from the latter by changing $x_0 \leftrightarrow y_0$ and $x \leftrightarrow y$, see Figure~\ref{F02}(b). Entanglement is now found whenever $y$ obeys inequalities (\ref{entcrit1}). 

%%%%%%%%%%%%%%%%%%%%%%%%%%%%%%%%%%
%%%%%%%%%%%%%%%%%%%%%%%%%%%%%%%%%%

The most striking thing about the above results is that all states $\rho_{W_k}$ share the same entanglement profile: they are entangled for $q \in (1/3, 1]$, regardless of $k$. 

Note that $\rho_{W_k}$ coincides with $\rho_B$ if $b_k = (3 q+1)/4$ and $b_j = (1-q)/4$, where $k = \operatorname{fixed}$ and $j \neq k$. That is, the states $\rho_{W_k}$ shown in Figure~\ref{F02} correspond to concrete convex combinations of the four Bell--states, where $\vert \beta_k \rangle$ plays a leading role.

\begin{figure}[h!]

\centering
\subfloat[][$\rho_B$, $\rho_S$, $s=1$]{\includegraphics[width=0.3\textwidth]{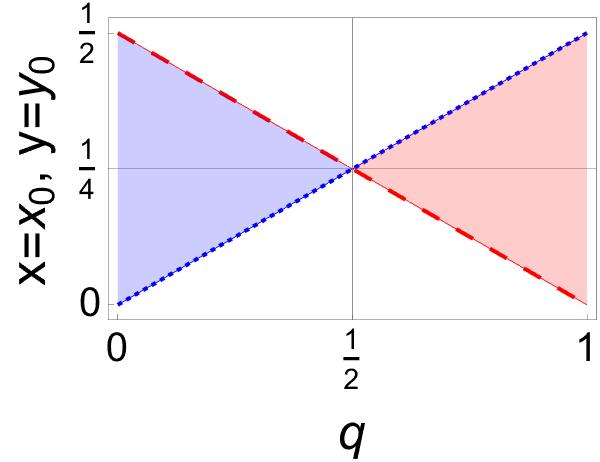}}
\hskip3ex
\subfloat[][$\rho_S$, $s=3/4$]{\includegraphics[width=0.3\textwidth]{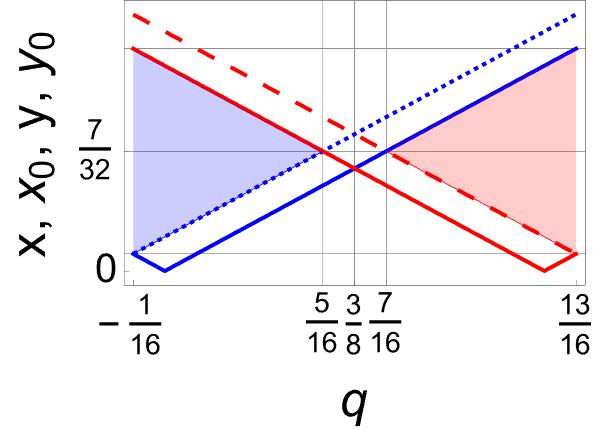}}
\hskip3ex
\subfloat[][$\rho_S$, $s=1/2$]{\includegraphics[width=0.3\textwidth]{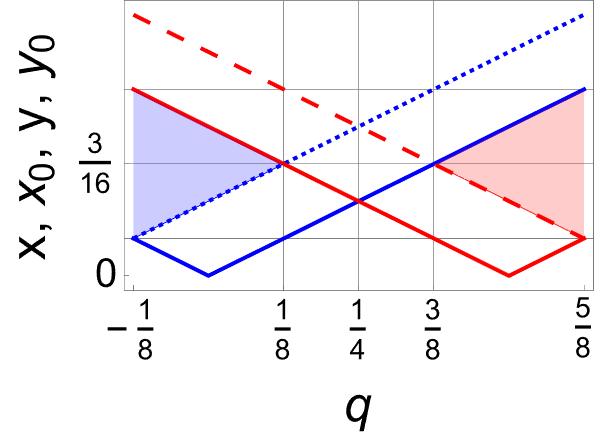}}
\vskip3ex
\subfloat[][$\rho_B$, $\rho_S$, $s=1$]{\includegraphics[width=0.3\textwidth]{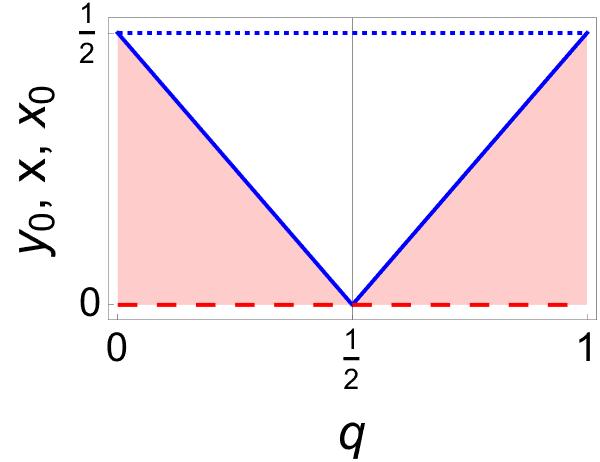}}
\hskip3ex
\subfloat[][$\rho_S$, $s=3/4$]{\includegraphics[width=0.3\textwidth]{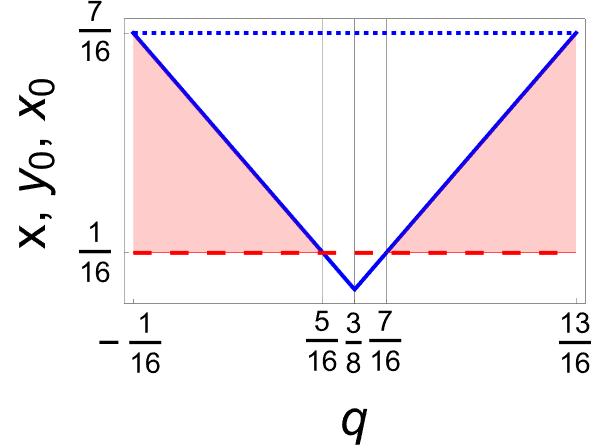}}
\hskip3ex
\subfloat[][$\rho_S$, $s=1/2$]{\includegraphics[width=0.3\textwidth]{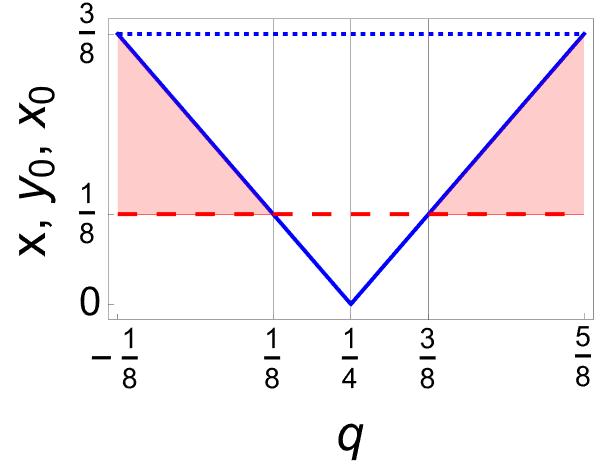}}

\caption{\footnotesize Entanglement conditions for the statistical mixture of Bell--states $\rho_B$ and the generalized Werner--state $\rho_S$, with only two nonzero coefficients $b_k =q$, $b_j = 1- q$, and $q_k=q$, $q_j = s-q$, respectively. In all cases $k$ and $j \neq k$ are fixed. The {\bf first row} refers to any pair $(k, j)$, except $(1,2)$ and $(3,4)$. The {\bf second row} illustrates the case $(1,2)$, for which $y=0$. The case $(3,4)$ is obtained from the second row after the changes $x_0 \leftrightarrow y_0$, $x \leftrightarrow y$. Color code follows the instructions in Figure~\ref{F02}.
}
\label{F03}
\end{figure}

To highlight the relevance of the Bell--states involved in any statistical mixture, let us analyze the (generalized) Werner--state (\ref{rhos}) with only two non-zero $q_k$ coefficients. If $s \neq 1$, this state also involves the four Bell--states. However, if $s=1$, it is reduced to a statistical mixture of only two Bell--states, those involved with coefficients $q_k \neq 0$. The model is illustrated in Figure~\ref{F03} with $b_k =q$, $b_j = 1- q$, and $q_k=q$, $q_j = s-q$, for $\rho_B$ and $\rho_S$ respectively. In all cases $k$ and $j \neq k$ are fixed. As in the previous case, what is common in all the states shown in Figure~\ref{F03} is their entanglement profile.

%%%%%%%%%%%%%%%%%%%%%%%%%%%%%%%%%%
%%%%%%%%%%%%%%%%%%%%%%%%%%%%%%%%%%

%---------------------------------------> Section
\subsection{Entanglement}
\label{entanglement}

In what follows we assume that $x_0$ and $y_0$ are known in advance. The description in terms of the point $p= (x,y) \in \mathcal S$ corresponds to determining the possible entanglement with respect to the coherences $r_{14}$ and $r_{23}$. 

The convex set $\mathcal S$ can be divided into regions (subsets) whose points define separable states and regions where entanglement may be found. We distinguish three main classes, described below.

\begin{figure}[h!]

\centering 
\includegraphics[width=0.25\textwidth]{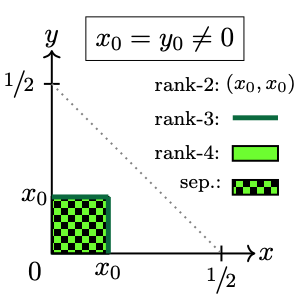}

\caption{\footnotesize There is {\bf no entanglement} if the populations of $\rho_X$ satisfy $r_{11} r_{44} = r_{22} r_{33} \neq 0$. In notation of the convex set $\mathcal S$, the latter means $x_{0} = y_{0} \neq 0$. The points $(x,y) \in \mathcal S$ that meet this condition form a square $\square_{x_0}$ of sides $x_0$, with $x_{0  \max} = 1/4$. 
}

\label{F04}
\end{figure}

%------------------------->
$\bullet$ Region $\mathcal M_0$. No entanglement is allowed if $r_{11} r_{44} = r_{22} r_{33} \neq 0$. That is, none of inequalities (\ref{entcrit2})-(\ref{entcrit1}) apply if $x_{0} = y_{0} \neq 0$. Instead we have $0 \leq x, y \leq x_0$, so we arrive at  a square $\square_{x_0}$ of sides $x_0$, with $x_{0  \max} = 1/4$, see Figure~\ref{F04}. These points produce separable mixed states $\rho_X$ that can be identified by their rank. According to the eigenvalues (\ref{lambdas}), the vertex $q_\star =(x_0 ,x_0)$ represents rank-2 states. In turn, $(x_0,y)$ and $(x,y_0)$, located respectively along the right-vertical and top sides of the square, define rank-3 states. Any other point of $\square_{x_0}$ gives rank-4 states.

%%%%%%%%%%%%%%%%%%%%%%%%%%%%%%%%%%
%%%%%%%%%%%%%%%%%%%%%%%%%%%%%%%%%%

As an example consider the states $\rho_B = \rho_S$ shown in Figure~\ref{F03}(a). Entanglement is lost precisely at $q=1/2$, where $x_0 = y_0 = 1/4$ puts vertex $q_{\star}= (x_0,x_0)$ to the middle of the hypotenuse of $\mathcal S$. The same is true for the states $\rho_{W_k}$ exhibited in Figure~\ref{F02}, at $q=0$.

\begin{figure}[htbp]
\centering
\subfloat[][Region $\mathcal M_1$]{\includegraphics[width=0.25\textwidth]{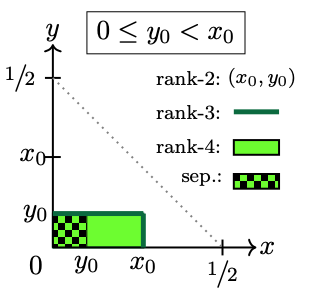}}
\hskip1cm
\subfloat[][$r_{22}=r_{33} =0$]{\includegraphics[width=0.25\textwidth]{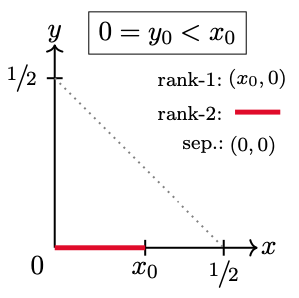}}

\caption{\footnotesize Entanglement conditions for states $\rho_{\mathcal M_1}$; the entanglement set is shaded-green, the chessboard region corresponds to separable states ({\bf a}) According to (\ref{entcrit2}), entanglement is encoded in a rectangle with base $x_0 - y_0$ and height $y_0$. The square $\square_{y_0}$ defines separable states ({\bf b}) Setting $y=y_0 =0$, with $r_{22}=r_{33} =0$, the states $\rho_{\mathcal M_x}$ acquire the form of the third matrix in Eq.~(\ref{otros}), which is  of rank-2 if $x< x_0$. The pure (rank-1) states $\vert \psi_{14} \rangle$ are achieved if $x= x_0$, which includes the Bell--states $\vert \beta_{1,2} \rangle$ when $x = x_{0 \max} =1/2$ 
}

\label{F05}
\end{figure}

%------------------------->
$\bullet$ Region $\mathcal M_1$. To satisfy inequalities (\ref{entcrit2}) it is necessary to take the ordering $0 \leq y_0 < x_0$, see for example the shaded-red area in Figures~\ref{F02} and \ref{F03}. The entanglement set is a rectangle lying on the horizontal leg of $\mathcal S$, as shown in Figure~\ref{F05}(a). The square $\square_{y_0}$ of separable states is located just to the left of such rectangle.

Using the eigenvalues (\ref{lambdas}) we find that vertex $q_0 = (x_0, y_0)$ gives rank-2 states. Points along the right-vertical side and the top of the rectangle, $(x_0,y)$ and $(x,y_0)$ respectively, yield rank-3 states. Any other point of this entanglement set provides rank-4 states.

%%%%%%%%%%%%%
%%%%%%%%%%%%%

A very important subset $\mathcal M_x \subset \mathcal M_1$, defined by  the coherence amplitudes $y = 0$ (regardless of $y_0$), provides points restricted to the horizontal leg of $\mathcal S$. Hereafter, the states defined by points $(x,0) \in \mathcal M_x$ will be denoted by $\rho_{\mathcal M_x}$. Since $y=0$, to find entanglement, our interest is addressed to the coherence amplitudes $x$ that satisfy $y_0 <x \leq x_0$. As an example consider the Werner--states $\rho_{W_{1,2}}$ illustrated in Figure~\ref{F02}(a), as well as states $\rho_S$ shown in the second row of Figure~\ref{F03}.

In general, by extending the domain of $x$ to the maximum we optimize the entanglement in $\mathcal M_x$. For this, it is enough to set $y_0=0$. We find three different \textcolor{red}{sets of} states 
\begin{equation}
\left(
\begin{array}{cccc}
r_{11} & 0 & 0 & x e^{i \theta}\\
0 & 0 & 0 & 0\\
0 & 0 & r_{33} & 0\\
x e^{-i\theta} & 0 & 0 & r_{44}
\end{array}
\right),
\quad
\left(
\begin{array}{cccc}
r_{11} & 0 & 0 & x e^{i \theta}\\
0 & r_{22} & 0 & 0\\
0 & 0 & 0 & 0\\
x e^{-i \theta} & 0 & 0 & r_{44}
\end{array}
\right),
\quad
\left(
\begin{array}{cccc}
r_{11} & 0 & 0 & x e^{i \theta}\\
0 & 0 & 0 & 0\\
0 & 0 & 0 & 0\\
x e^{-i \theta} & 0 & 0 & r_{44}
\end{array}
\right).
\label{otros}
\end{equation}
Neither of the first two matrices in (\ref{otros}) satisfy the factorization of their elements as required in Proposition~Q, so they represent mixed states of rank-3 or rank-2, see Eq.~(\ref{rankA}). In turn, if the elements of the third matrix are factorized we have $x = x_0 = \vert \alpha_1 \alpha_4 \vert$, so it represents the pure (rank-1) state $\vert \psi_{14} \rangle$ defined in (\ref{res1}). In particular, if $x= x_{0 \max} =1/2$ we recover the Bell-states $\vert \beta_1 \rangle$ and $\vert \beta_2 \rangle$. If their elements cannot be factorized, this matrix represents a mixed state of rank-2, see Figure~\ref{F05}(b). 

%%%%%%%%%%%%%
\begin{figure}[htbp]
\centering
\subfloat[][Region $\mathcal M_2$]{\includegraphics[width=0.25\textwidth]{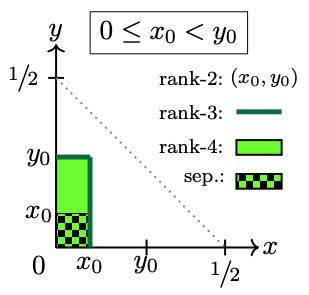}}
\hskip2cm
\subfloat[][$r_{11}=r_{44} =0$]{\includegraphics[width=0.25\textwidth]{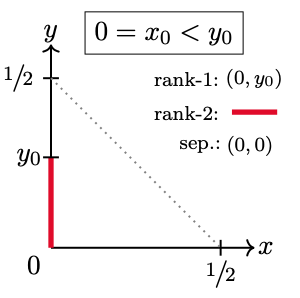}}

\caption{\footnotesize Entanglement conditions for states $\rho_{\mathcal M_2}$ ({\bf a}) According to (\ref{entcrit1}), we find entanglement in a rectangle of base $x_0$ and height $y_0-x_0$. The square $\square_{x_0}$ gives separable states ({\bf b}) If $x=x_0 =0$, with $r_{11}=r_{44} =0$, the states $\rho_{\mathcal M_y}$ acquire the form of the third matrix in Eq.~(\ref{forms1}), which is of rank-2 if $y< y_0$. The pure (rank-1) states $\vert \psi_{23} \rangle$ are achieved if $y= y_0$, which includes the Bell--states $\vert \beta_{3,4} \rangle$ when $y = y_{0 \max} =1/2$. 
}

\label{F06}
\end{figure}
%%%%%%%%%%%%%

%------------------------->
$\bullet$ Region $\mathcal M_2$. Inequalities (\ref{entcrit1}) are satisfied by imposing the ordering $0 \leq x_{0} < y_{0}$, see for example the shaded-blue area in Figures~\ref{F02} and \ref{F03}. The entanglement subset is also a rectangle, but in this case the base is placed just above the separable-states square $\square_{x_0}$, see Figure~\ref{F06}(a). Points along the right-vertical side and the top of the rectangle, respectively $(x_0,y)$ and $(x,y_0)$, give rise to rank-3 states. In turn, vertex $q_0 =(x_{0},y_{0})$ corresponds to rank-2 states. Any other point provides rank-4 states.

In this case the subset $\mathcal M_y \subset \mathcal M_2$, defined by the coherence amplitudes $x = 0$, provides points restricted to the vertical leg of $\mathcal S$ and defines states $\rho_{\mathcal M_y}$. Our interest now turns to the region $x_0 < y \leq y_0$. As an example consider the Werner--states $\rho_{W_{3,4}}$ exhibited in Figure~\ref{F02}(b) and the $y$-version of the states $\rho_S$ shown in the second row of Figure~\ref{F03}. The optimization is achieved by setting $x_0 =0$, which results in the following states
\begin{equation}
\left(
\begin{array}{cccc}
0 & 0 & 0 & 0\\
0 & r_{22} & y e^{i \phi} & 0\\
0 & y e^{-i \phi} & r_{33} & 0\\
0 & 0 & 0 & r_{44}
\end{array}
\right),
\quad
\left(
\begin{array}{cccc}
r_{11} & 0 & 0 & 0\\
0 & r_{22} & y e^{i \phi} & 0\\
0 & y e^{-i \phi} & r_{33} & 0\\
0 & 0 & 0 & 0
\end{array}
\right),
\quad
\left(
\begin{array}{cccc}
0 & 0 & 0 & 0\\
0 & r_{22} & y e^{i \phi} & 0\\
0 & y e^{-i \phi} & r_{33} & 0\\
0 & 0 & 0 & 0
\end{array}
\right).
\label{forms1}
\end{equation}
The first two matrices do not satisfy Proposition~Q, so they represent mixed states of rank-3 or rank-2, see Eq.~(\ref{rankB}). If the elements of the third matrix are factorized we have $y = y_0 = \vert \alpha_2 \alpha_3 \vert$, so it represents the pure (rank-1) state $\vert \psi_{23} \rangle$ defined in (\ref{res2}). If $y= y_{0 \max} =1/2$, we recover the Bell-states $\vert \beta_3 \rangle$ and $\vert \beta_4 \rangle$. If their elements cannot be factorized, this matrix is a mixed state of rank-2, see Figure~\ref{F06}(b). 

The classification developed above is very close to the parametrization of two-qubit X--density matrices with a fixed rank reported in Section~2.1 of \cite{Men14}. The latter, based on the coefficients that define the characteristic equation for the density operator, considers the Newton-Girard formulae \cite{Ber09} and a very concrete parametrization of the matrix-elements of $\rho_X$. Beyond using the computational basis, our treatment is free of parametrization for the matrix elements $r_{jk}$, so it aims to be more practical and, therefore, universal. Nevertheless, the parameters used in \cite{Men14} are very valuable as they comprise the information needed to search for a unitary transformation that converts arbitrary two-qubit states into their X counterparts.

%---------------------------------------> Section
\subsection{Geometric $L$-measure}
\label{measures}

In the previous sections we have identified the regions of the convex set $\mathcal S$ whose points would imply entanglement in state $\rho_X$. We know, for example, that squares $\square_{x_0}$ and $\square_{y_0}$ must be discarded because all their points represent separable states. However, we have not quantified the amount of entanglement that distinguishes some points from others in any other region, nor have we analyzed which region offers better entanglement conditions. To address these points we proceed as follows.

We seek to ensure that entangled states ``resemble'' separable states as little as possible. Translated into the language of the convex set $\mathcal S$, this statement means that if $p$ encodes entanglement then it should be as far away as possible from $\square_{x_0}$  and $\square_{y_0}$. The further remote, the greater the entanglement. With this in mind, consider the metric $d (\vec r_1, \vec r_2) =  \max \{ \vert x_2 - x_1\vert, \vert y_2 -y_1 \vert \}$, where $\vec r_k = (x_k,y_k) \in \mathbb R^2$, $k=1,2$. We now introduce a geometric measure of entanglement for two-qubit systems. 

{\bf  Definition.} ($L$-measure of entanglement) Given the extreme points $x_0$ and $y_0$, let $\mathcal S$ be the convex set (\ref{convex}) and $\square_{z_0} \subset \mathcal S$ be the subset of points defining separable states. The $L$-measure of entanglement, applied to the point $p \in \mathcal S$, is the distance between $p$ and the closest point $q \in \square_{z_0}$ given by
\[
L = \xi_0 \inf_q d(q,p),
\]
with $\xi_0$ a normalization factor.

%%%%%%%%%%%%%%%%%%%%%%%%%%%%%%%%%%
\begin{figure}[h!]
 
\centering
\subfloat[][Region $\mathcal M_1$]{\includegraphics[width=0.17\textwidth]{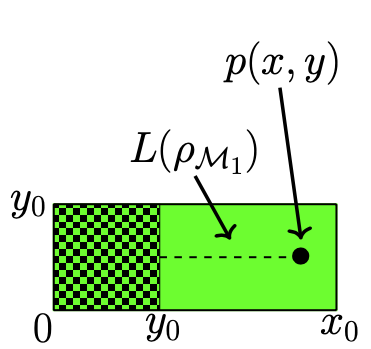}}
\hskip2cm
\subfloat[][Region $\mathcal M_2$]{\includegraphics[width=0.17\textwidth]{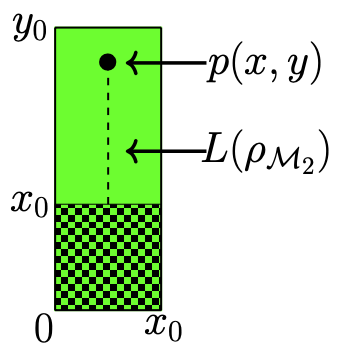}}

\caption{\footnotesize Illustration of the geometric notion of distance used to define the measure~$L$ for states $\rho_X$ defined by points in regions $\mathcal M_1$ or $\mathcal M_2$.
}

\label{F07}
\end{figure}
%%%%%%%%%%%%%%%%%%%%%%%%%%%%%%%%%%

Let us apply the measure $L$ to the points $p \in \mathcal S$ that define the states classified in the previous section. 

States $\rho_{\mathcal M_1}$ are such that $\inf d(q,p)$ is given by the points $q=(y_0,y)$, see Figure~\ref{F07}(a). Therefore
\be
\left. L (\rho_{\mathcal M_1}) \right\vert_{q=(y_0, y)} =  2 \max \left\{0, x-y_0 \right\} = \left\{
\begin{array}{cc}
0 & \operatorname{if} \, x< y_0\\[.5ex]
2 (x- y_0) & \operatorname{if} \, x > y_0
\end{array}
\right.
\label{ele5}
\ee
Note that $x=x_0$ defines the furthest point from the zone of separable states, therefore $L_{\max} (\rho_{\mathcal M_1}) = 2(x_0 -y_0)$. That is, among these states, the highest value of the $L$-measure is provided by the rank-3 states associated with the points $(x_0,y)$.

For states $\rho_{\mathcal M_2}$ we find that $q=(x,x_0)$ defines $\inf d(q,p)$, see Figure~\ref{F07}(b). Then
\be
\left. L (\rho_{\mathcal M_2}) \right\vert_{q=(x,x_0)}= 2 \max \left\{ 0, y-x_0 \right\} 
= \left\{
\begin{array}{cc}
0 & \operatorname{if} \, y< x_0\\[.5ex]
2 (y- x_0) & \operatorname{if} \, y > x_0
\end{array}
\right.
\label{ele3}
\ee
In this case the highest value of the $L$-measure is provided by the rank-3 states associated with the points $(x, y_0)$, so that $L_{\max} (\rho_{\mathcal M_2})  = 2(y_0-x_0)$.

Combining the previous results we obtain the general rule
\be
\begin{array}{rl}
L(\rho_X)  & =  2 \max \left\{ 0, x-y_0, y-x_0 \right\} \\[1ex]
& = 
2 \max \left\{0, \vert r_{14}\vert - \sqrt{r_{22} r_{33}}, \, \vert r_{23}\vert - \sqrt{r_{11} r_{44}} \, \right\}.
\end{array}
\label{ele}
\ee
To find entanglement it is enough for one of the subtractions of (\ref{ele}) to be positive.

As for systems composed of subsystems with equal entropy, see Section~\ref{equal}, we have immediate results:
\bea
L(\rho_B) = 2 \max \left\{ 0, \frac{\vert b_1 -b_2 \vert}{2} - \frac{(b_3+b_4)}{2}, \frac{\vert b_3 -b_4 \vert}{2} - \frac{(b_1+b_2)}{2}
\right\},
\label{medidas1}\\[1ex]
L(\rho_{W_k}) = 2 \max \left\{ 0, \frac{\vert q \vert}{2} - \frac{\vert 1 -q \vert}{4} \right\},
\label{medidas2}\\[1ex]
L(\rho_S) = 2 \max \left\{ 0, \frac{\vert q_1 -q_2\vert}{2} - \varrho_{22}, \frac{\vert q_3 -q_4\vert}{2} - \varrho_{11}
\right\}.
\label{medidas3}
\eea

To illustrate the applicability of the above expressions let us consider the state (\ref{rhobm2})-(\ref{rhobm3}). From (\ref{medidas1}) one has
\be
\left. L(\rho_B) \right\vert_{b_3=b_4=\kappa}= 2 \max \left\{ 0, \frac{\vert b_1 -b_2 \vert}{2} - \kappa \right\}.
\ee
Therefore, $\left. \rho_B \right\vert_{b_3=b_4=\kappa}$ is disentangled if $\vert b_1 - b_2 \vert \leq 2 \kappa = 1- (b_1+b_2)$. For these parameters the interference between $\vert e_1 \rangle = \vert 00 \rangle$ and $\vert e_4 \rangle = \vert 11 \rangle$ is so small that the system ends up behaving like a statistical mixture (a classical state) of the base states. On the other hand, for $\vert b_1 - b_2 \vert > 2 \kappa$ we find entanglement with measure $\left. L(\rho_B) \right\vert_{b_3=b_4=\kappa}= \vert b_1 - b_2\vert - 2 \kappa$
. The extreme condition $\kappa=0$ shows that the statistical mixture of the Bell--states $\vert \beta_1 \rangle$ and $\vert \beta_2 \rangle$ is entangled, with a measure $L$ that is greater as the interference between $\vert 00 \rangle$ and $\vert 11 \rangle$ is more intense. In particular, setting $b_1 =q$ and $b_2 = 1-q$ we have $\left. L(\rho_B) \right\vert_{b_3=b_4=0}= \vert 2q -1 \vert$. So state $\left. \rho_B \right\vert_{b_3=b_4=0}$ is disentangled for $q=1/2$ ($b_1=b_2$), as it is shown in Figures~\ref{F03}(a) and \ref{F03}(d).

In general, if $x$ and $y$ take the maximum value in their respective domains we have
\[
L_{\max} (\rho_X)  =  2 \max \left\{ 0, x_0-y_0, y_0-x_0 \right\} = 
2 \vert x_0 - y_0 \vert =
2 \left\vert \sqrt{r_{11} r_{44}} - \sqrt{r_{22} r_{33} } \, \right\vert.
\]
That is, when $x_0$ and $y_0$ are nonzero and different from each other, the strongest entanglement is exhibited by  rank-3 states $\rho_X$. On the other hand, if $x_0 =0$ or $y_0=0$, the $L$-measure identifies rank-1 (pure) states as those with greater entanglement. In particular, if $x_0 = x_{0 \max} =1/2$ or $y_0 =y_{0 \max}= 1/2$, such states are not only pure but coincide with one of the Bell--states.

The most notable of the previous results is that the mathematical expression of the geometric measure~$L$ coincides precisely with the Hill-Wootters concurrence calculated for the $\rho_X$ states \cite{Wan06,Yu07,Que12}. In fact, according with \cite{Hil97,Woo98}, the concurrence for any state $\rho$ is given by $C(\rho) = \max \lbrace 0, \, \sqrt{\ell_1} - \sqrt{\ell_2} - \sqrt{\ell_3} - \sqrt{\ell_4} \rbrace$. The $\ell_k$'s are the eigenvalues (in decreasing order) of the matrix $R_{\rho} = \rho \left( \sigma_2 \otimes \sigma_2 \right) \rho^{*} \left( \sigma_2 \otimes \sigma_2 \right)$, with $\sigma_2$ the well-known anti-diagonal Pauli-matrix.

Using state $\rho_X$ to construct $R_{\rho}$, a simple calculation produces the eigenvalues $(x_0 \pm x)^2$ and $(y_0 \pm y)^2$, where no order has yet been established. From (\ref{rhocond}) we know that the quantities in parentheses are not negative, so we have two different ways of sorting $\sqrt{\ell_k }$ in descending order, with $y_0+y$ or $x_0 +x$ leading. A little algebra produces the result
\[
C(\rho_X) = \left\{
\begin{array}{rl}
2 \max \left\{ 0, y-x_0 \right\} , & y_0 +y> x_0 + x\\[1ex]
2 \max \left\{ 0,  x-y_0 \right\}, &  x_0 + x > y_0 +y
\end{array}
\right. 
\]
which once summarized is written as follows \cite{Wan06,Yu07,Que12}:
\be
C(\rho_X) = 2 \max \left\{0, \vert r_{14}\vert - \sqrt{r_{22} r_{33}}, \, \vert r_{23}\vert - \sqrt{r_{11} r_{44}} \, \right\}.
\label{comun}
\ee
Comparing (\ref{comun}) with (\ref{ele}) we find  $C(\rho_X) = L(\rho_X)$. 

That is, the Hill-Wootters concurrence coincides with our $L$-measure when applied to state $\rho_X$. With this result we have given $C(\rho_X)$ a geometric meaning that is not obvious without the structure of the convex set $\mathcal S$. What is even better is that, unlike concurrence, the geometric measure distinguishes the rank of states with maximum entanglement.

In this context, the $L$-measure of Werner--states (\ref{medidas2}) deserves special attention. First of all, this verifies that all states $\rho_{W_k}$ share the same entanglement profile: they are entangled for $q \in (1/3, 1]$, regardless of $k$, see Figure~\ref{F02}. On the other hand, two-qubit Werner--states can be prepared experimentally in such a way that the related hierarchy of quantum correlations can also be demonstrated. Indeed, in \cite{Jir21,Abo23}, states (\ref{werner}) are prepared for various values of the mixing parameter $q$, so that concurrence $C(\rho_{W_k}) =L({W_k}) $ is measured with good agreement of (\ref{medidas2}).

Once we have proven the identity $C=L$, it is appropriate to construct a geometric version of the entanglement of formation \cite{Ben96}. Following \cite{Woo98} we obtain
\be
\varepsilon (\rho_{\mathcal M_1}) = \left\{
\begin{array}{cc}
0 & \operatorname{if}\,  x < y_0\\[.5ex]
h \left( \frac{ 1 + \sqrt{1 - 4 (x-y_0)^2 }}{2}
\right) &  \operatorname{if}\,  x > y_0
\end{array}
\right.
\label{E1}
\ee
and 
\be
\varepsilon (\rho_{\mathcal M_2}) = \left\{
\begin{array}{cc}
0 & \operatorname{if}\,  y < x_0\\[.5ex]
h \left( \frac{ 1 + \sqrt{1 - 4 (y-x_0)^2 }}{2}
\right) &  \operatorname{if}\,  y > x_0
\end{array}
\right.
\label{E2}
\ee
where $h(z) = -z \log_2 z - (1-z) \log_2 (1-z)$ is the Sahnnon binary entropy function.

%---------------------------------------> Section
\subsection{Determination of entanglement}

\begin{figure}[h!]

\centering 
\subfloat[][Werner states]{\includegraphics[width=0.3\textwidth]{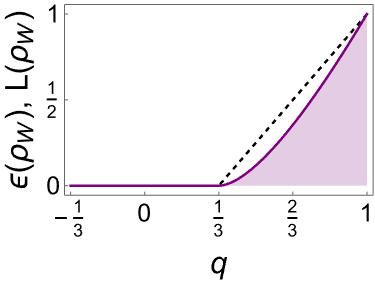}}
\hskip1cm
\subfloat[][Two-Bell mixed states]{\includegraphics[width=0.3\textwidth]{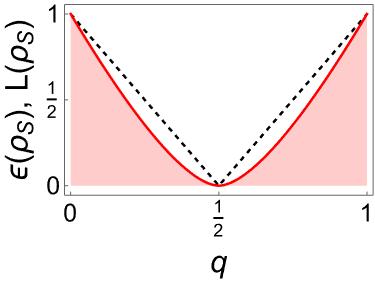}}

\caption{\footnotesize The geometric measure~$L$ (dotted curve) and entanglement of formation $\varepsilon$ (continuous curve) applied to ({\bf a}) the Werner--states $\rho_{W_k}$ and ({\bf b}) the statistical mixture $\rho_B$ of only two Bell--states. There is no entanglement for half the domain of $q$ in $\rho_{W_k}$, compare with Figure~\ref{F02}. In turn, states $\rho_B$ lack entanglement only at one point ($q = 1/2$) of the corresponding domain, compare with Figures~\ref{F03}(a) and~\ref{F03}(d).
}

\label{F08}
\end{figure}

The entanglement of the Werner--states $\rho_{W_k}$, quantified by $L$ and $\varepsilon$, is shown in Figure~\ref{F08}(a) (see also Figure~\ref{F02}). The optimization is achieved in half of the domain of $q$ only. The experimental realization of Werner--states and the measurement of their entanglement can be found in \cite{Jir21,Abo23}.

%%%%%%%%%%%%%%%%%%%%%%%%%%%%%%%%%%
%%%%%%%%%%%%%%%%%%%%%%%%%%%%%%%%%%

Figure~\ref{F08}(b) shows the result of applying $L$ and $\varepsilon$ to the convex combinations \textcolor{red}{$\rho_B$} of Bell--states exhibited in Figures~\ref{F03}(a) and \ref{F03}(d). This time the optimization is over almost the entire domain of $q$, mainly in the neighborhoods of $q = 0$ and $q = 1$, which serve as accumulation points. These systems are separable only for $q =1/2$. 

\begin{figure}[h!]

\centering 
\subfloat[][$s=3/4$]{\includegraphics[width=0.3\textwidth]{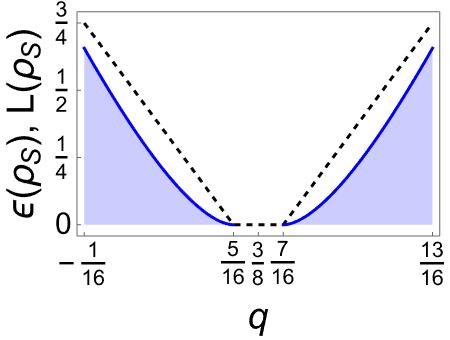}}
\hskip1cm
\subfloat[][$s=1/2$]{\includegraphics[width=0.3\textwidth]{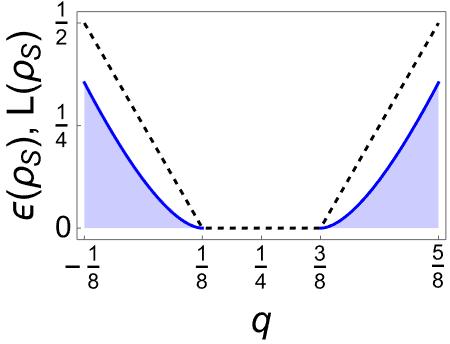}}

\caption{\footnotesize The geometric measure~$L$ (dotted curve) and entanglement of formation $\varepsilon$ (continuous curve) applied to the generalized Werner--states shown in Figure~\ref{F03}. The case depicted in Figure~\ref{F08}(b) is recovered if $s=1$. In general, the maximum values of $L$ and $\varepsilon$ decrease with $s$ while the region where they become zero increases.
}

\label{F09}
\end{figure}

On the other hand, applying $L$ and $\varepsilon$ to the generalized Werner--states shown in Figure~\ref{F03} produces results that depend on $s$, as expected. In Figure~\ref{F09} we see that the maximum of $L$ and $\varepsilon$ decreases with $s<1$. Furthermore, the interval of $q$ for which $L$ is equal to zero becomes increasingly larger. Thus, any perturbation that modifies the original statistical mixture (i.e., the value $s=1$) drastically affects entanglement: as mixing is maximized, the system loses quantum properties, making it increasingly classical.

%%%%%%%%%%%%%%%%%%%%%%%%%%%%%%%%%%
%%%%%%%%%%%%%%%%%%%%%%%%%%%%%%%%%%

Another way to investigate the transition from entangled to separable configurations of a given state $\rho$ is to use the concept of robustness \cite{Vid99}, which quantifies entanglement with respect to the maximally mixed state $\rho_{\star}$ and may be expressed in terms of the concurrence \cite{Kry21}. Referring to the optimized states of two qubits derived in the previous sections, the main point is to consider the density operator
\be
\widetilde\rho_X= (1-\omega) \rho_X + \omega \rho_{\star}, \quad w \in [0,1].
\ee
The robustness of the optimized $\rho_X$ corresponds to the minimal value of $\omega$, written $\omega_0$, for which $\widetilde\rho_X$ becomes disentangled \cite{Vid99,Kry21}. The most important feature of $\widetilde\rho_X$ is that it is also X--shaped. Therefore, the $L$--measure (\ref{ele}) is automatically applicable. A simple calculation yields
\bea
L(\widetilde\rho_B) = 2 \max \left\{ 0, (1-\omega) L_1(\rho_B) - \frac{\omega}{4}, (1-\omega) L_2(\rho_B) - \frac{\omega}{4}
\right\},
\label{robus1}\\[1ex]
L(\widetilde\rho_{W_k}) = 2 \max \left\{ 0, (1-\omega) L_1(\rho_{W_k}) - \frac{\omega}{4} \right\},
\label{robus2}\\[1ex]
L(\widetilde\rho_S) = 2 \max \left\{ 0, (1-\omega) L_1(\rho_S) - \frac{\omega}{4}, (1-\omega) L_2(\rho_S) - \frac{\omega}{4}
\right\},
\label{robus3}
\eea
where $L_j(\rho_X)$ stands for the first ($j=1$) and second ($j=2$) subtractions occurring in the $L$-measures (\ref{medidas1})-(\ref{medidas3}). Assuming that subtraction $L_j(\rho_X) >0$ defines the value of $L(\rho_X)$, we obtain the robustness of the optimized states of two qubits:
\be
\omega_0 = \frac{L_j(\rho_X)}{L_j (\rho_X) + 1/4}.
\ee
For the Werner--states shown in Figure~\ref{F08}(a), this expression takes the maximal value 4/5 at $q=1$, and minimal value 0 if $q \leq 1/3$. These values correspond to the maximally entangled expression of $\rho_{W_k}$ and a completely disentangled form of $\rho_{W_k}$, respectively. Similar conclusions are obtained for the two-Bell mixed states shown in Figure~\ref{F08}(b). In turn, the robustness of the generalized Werner--states shown in Figures~\ref{F09}(a) and \ref{F09}(b) take the maximal values 3/4 and 2/3, respectively. 

%---------------------------------------> Section
\subsection{Time-dependent systems}

To further extend the applicability of our treatment, we now study what happens to entanglement when the system is explicitly time-dependent. 

Suppose we have at hand a pair of two-level atoms which are entangled in energy, say $\vert \psi_{\operatorname{2at}} \rangle = \vert \beta_3 \rangle$. Let us place these atoms in independent, isolated and identical electromagnetic cavities, each of which containing exactly $n$ photons. The state of the entire system can be written as $\vert \Phi \rangle = \tfrac{1}{\sqrt 2} \left( \vert +, n \rangle \otimes \vert -, n \rangle + \vert -, n \rangle \otimes \vert +, n \rangle \right)$, where $\vert n \rangle$ represents the state of the quantized field contained in each cavity. Besides $\vert + \rangle \equiv \vert 0 \rangle$ and $\vert - \rangle \equiv \vert 1 \rangle$. Following \cite{Enr14,Qui15,Qui16}, with $\gamma$ a coupling factor, the time-evolved state acquires the form
\[
\begin{array}{rl}
\vert \Phi (t) \rangle  & = \cos ( \gamma t \sqrt{n+1} \/) \left[ \cos ( \gamma t \sqrt{n} \/ ) \vert \varphi_1 \rangle
-i \sin ( \gamma t \sqrt{n} \/ ) \vert \varphi_2 \rangle \right]\\[1.5ex]
& \quad -i \sin ( \gamma t \sqrt{n+1} \/) \left[ \cos ( \gamma t \sqrt{n} \/) \vert \varphi_3 \rangle
-i \sin ( \gamma t \sqrt{n} \/) \vert \varphi_4 \rangle \right],
\end{array}
\]
with $ \vert \varphi_1 \rangle = \vert \Phi \rangle_{t=0}$, and 
\[
\begin{array}{l}
\vert \varphi_2 \rangle = \tfrac{1}{\sqrt 2} \left(  \vert +, n \rangle \otimes \vert +, n-1 \rangle + \vert +, n-1 \rangle \otimes \vert +, n \rangle \right),\\[1.5ex]
\vert \varphi_3 \rangle = \tfrac{1}{\sqrt 2} \left(  \vert -, n+1 \rangle \otimes \vert -, n \rangle + \vert -, n \rangle \otimes \vert -, n+1 \rangle \right),\\[1.5ex]
\vert \varphi_4 \rangle = \tfrac{1}{\sqrt 2} \left(  \vert -, n+1 \rangle \otimes \vert +, n-1 \rangle + \vert +, n-1 \rangle \otimes \vert -, n+1 \rangle \right).
\end{array}
\]
The reduced time-dependent diatomic state results 
\be
\rho_{\operatorname{2at}} (t)=  \operatorname{Tr}_{\operatorname{fields}} \rho(t) =
\left(
\begin{array}{cccc}
r_{11} (t) & 0 & 0 & 0\\[.5ex]
0 & y_0 (t) & y (t) & 0\\[.5ex]
0 & y (t) & y_0 (t) & 0\\[.5ex]
0 & 0 & 0 & r_{44} (t)
\end{array}
\right),
\label{atom}
\ee
where
\[
y_0 (t) = \tfrac12 \cos^2 \left( \gamma t \sqrt{n+1} \right) \cos^2 \left( \gamma t \sqrt{n} \right) + \tfrac12 \sin^2 \left( \gamma t \sqrt{n+1} \right) \sin^2 \left( \gamma t \sqrt{n} \right)
\]
and 
\[
y (t) = \tfrac12 \cos^2 \left( \gamma t \sqrt{n+1} \right) \cos^2 \left( \gamma t \sqrt{n} \right).
\]
In turn, the extreme point $x_0 = \sqrt{r_{11} r_{44} }$ is obtained from the matrix-elements
\[
r_{11} (t) =  \cos^2 \left( \gamma t \sqrt{n+1} \right) \sin^2 \left( \gamma t \sqrt{n} \right), \,\,
r_{44} (t) =  \sin^2 \left( \gamma t \sqrt{n+1} \right) \cos^2 \left( \gamma t \sqrt{n} \right).
\]

%%%%%%%%%%%%%%%%%%%%%%%%%%%%%%%%%%
\begin{figure}[h!]

\centering 
\subfloat[][]{\includegraphics[width=0.3\textwidth]{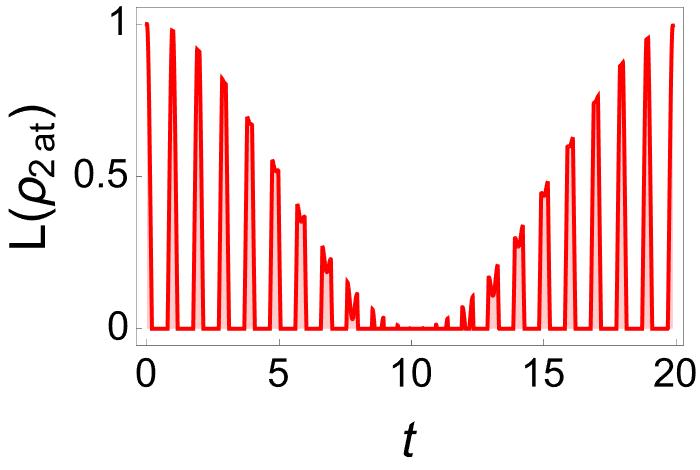}}
\hskip1cm
\subfloat[][]{\includegraphics[width=0.3\textwidth]{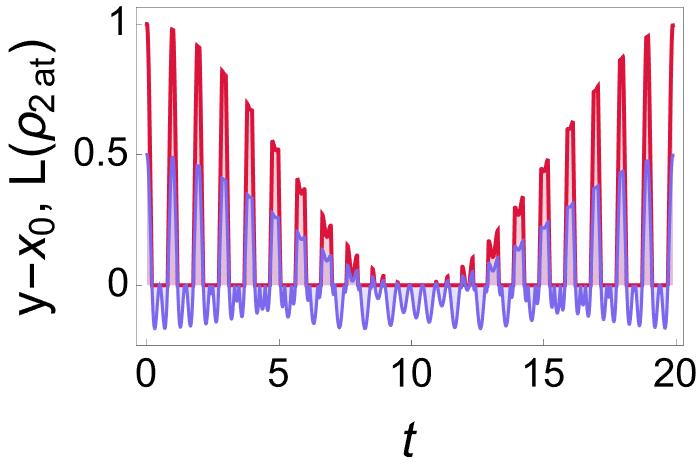}}

\caption{\footnotesize ({\bf a}) The geometric measure $L$ applied to the time-dependent diatomic state $\rho_{\operatorname{2at}}$ for $\gamma =1$ and $n=10$.  As time passes, $L$ decreases from 1 to zero and then increases until it is almost equal to 1. This process is repeated over and over again ({\bf b}) Comparison of  $L(\rho_{\operatorname{2at}})$ with the difference $y-x_0$. Clearly, inequality $y> x_0$ is violated in the same time intervals where $L=0$.
}

\label{F10}
\end{figure}
%%%%%%%%%%%%%%%%%%%%%%%%%%%%%%%%%%

The profile of the diatomic state $\rho_{\operatorname{2at}}$ fits the structure of $\rho_{\mathcal M_y}$ at any time. The $L$-measure returns the value 1 at $t=0$, where the diatomic system is in the Bell--state $\vert \beta_3 \rangle$. As time passes, $L$ decreases to zero and then increases until it is almost equal to 1. This process is repeated over and over again, see Figure~\ref{F10}(a). The time intervals where entanglement is zero coincide with the violation of inequality $x_0 < y$, see Figure~\ref{F10}(b).

\begin{figure}[h!]

\centering 
\subfloat[][]{\includegraphics[width=0.3\textwidth]{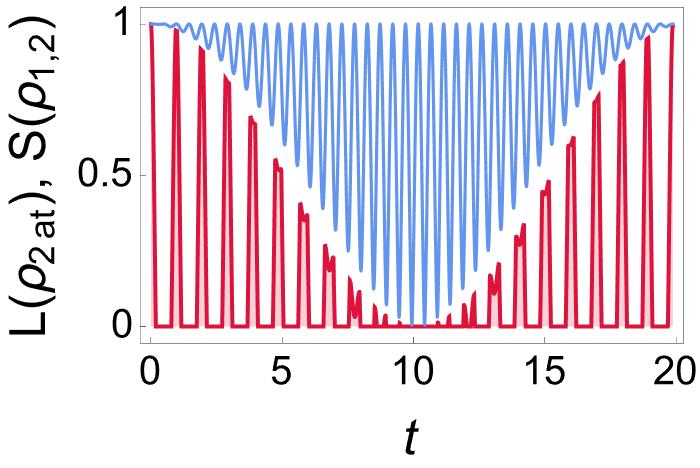}}
\hskip2cm
\subfloat[][]{\includegraphics[width=0.3\textwidth]{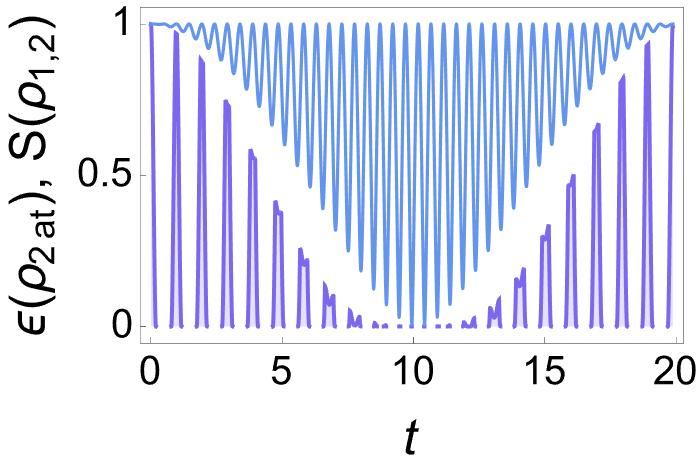}}

\caption{\footnotesize The envelope formed by the minima of the one-qubit entropy (blue curve) defines an upper bound for the entanglement measures of the time-dependent diatomic state $\rho_{\operatorname{2at}}$ ({\bf a}) The geometric measure $L$  is red-shaded  ({\bf b}) The entanglement of formation $\epsilon$ is purple-shaded.
}

\label{F11}
\end{figure}

We now focus our attention on the reduced states of $\rho_{\operatorname{2at}}$, which have the same entropy $S(\rho_1) = S(\rho_2)$. In Figure~\ref{F11} we have depicted the result of applying $L$ and $\varepsilon$ to $\rho_{\operatorname{2at}}$, together with the entropy $S(\rho_{1,2})$. We see that $S$ reaches its maximum value at different times. Some of such moments define also the local maxima of $L$ and $\varepsilon$, but others are in time intervals where there is no entanglement. As it is well known, the latter means that finding reduced states of maximum mixing does not provide information about the entanglement (if any) of the entire system if it is in a mixed state. However, the minima of $S$ play a different role in the system we are dealing with.

The envelope formed by the minima of S establishes an upper limit for the entanglement measure. To our knowledge, this surprising property of entropy shared by reduced states has not been reported in the literature on the matter.

The phenomenon described above is also present when we use any other Bell--state as initial condition instead of $\vert \beta_3 \rangle$. If we start from $\vert \beta_4 \rangle$ the structure of $\rho_{\operatorname{2at}}$ is similar to the one we find in (\ref{atom}). But starting from $\vert \beta_{1,2} \rangle$ we arrive at a diatomic state of the form $\rho_{\mathcal M_x}$. In any case, what we find is that the envelope formed by the minima of $S$ represents an upper bound on the entanglement strength as time progresses.

%%%%%%%%%%%%%%%%%%%%%%%%%%%%%%%%%%
%%%%%%%%%%%%%%%%%%%%%%%%%%%%%%%%%%

%---------------------------------------> Section
\section{Conclusions}
\label{conclu}

We have established the necessary conditions to study entanglement in two-qubit systems by using the minimum of essential parameters. 

The idea has been to construct density operators whose reduced one-qubit states share the same entropy, regardless of whether the state of the entire system is pure or mixed. The latter leads to the identification of $\rho_X$ states that have the capacity to host pairs of identical populations while reducing the number of coherences involved.

With the above results we have constructed a convex representation that facilitates the study of entanglement in two-qubit systems. The description is purely geometric and is based on two of the main characteristics of the X--states:

i) To obtain positive semidefinite operators $\rho_X$, the coherence amplitudes $x = \vert r_{14} \vert$ and $y = \vert r_{23} \vert$ should be bounded from above by the corresponding populations, $x_0 = \sqrt{r_{11} r_{44} }$ and $y_0 = \sqrt{r_{22} r_{33} }$. Otherwise, $\rho_X$ is not a quantum state.

ii) To find entanglement, $x$ and $y$ should be bounded from below by the  complementary populations, $y_0$ and $x_0$, respectively. Otherwise, state $\rho_X$ is separable.

Taking full advantage of such representation we have introduced a geometric measure $L$, which is defined as the distance between $p$ (the point under study) and the closest point $q$ that defines separable states. The more remote, the greater the entanglement.

The $L$-measure reproduces the results of the Hill--Wootters generalized concurrence $C$ for states $\rho_X$. That is, our results give $C$ a geometric meaning that is not obvious without the convex structure introduced in this work. What is even better is that, unlike $C$, the geometric measure $L$ distinguishes the rank of states with maximum entanglement.

We have found that rank-3 states $\rho_X$ exhibit the strongest entanglement when $x_0$ and $y_0$ are nonzero and different from each other. These states have no locally producible quantum correlations since no rank-3 or rank-4 state can be produced by local operations of one of the parties on a classically correlated state \cite{Ges12,Man15}. If $x_0 =0$ and $y=y_0$, or $y_0 =0$ and $x=x_0$, states $\rho_X$ are pure (rank-1) and yield strong entanglement. The Bell--states fall into this category by maximizing to 1 the result given by the measure~$L$. It seems that states of higher ranks are more useful for quantum information procedures \cite{Man15}. In particular, states of rank-3 can be used to reconstruct pure states by a remote party \cite{Wan13}. In turn, rank-4 states (which inhabit the different entanglement regions of $\mathcal S$) are useful for reconstructing arbitrary states \cite{Man15}.

Note that the universality of two-qubit X–states ensures the utility of our geometric representation for studying entanglement in any configuration of two-qubit states. Indeed, it has been shown that `for every two-qubit state there is a two-qubit X--state of same spectrum and entanglement, as measured by concurrence, negativity or relative entropy of entanglement' \cite{Men14}. In this sense, no entanglement is lost by avoiding certain coherences of the initial state $\rho$ during the optimization process that we have developed throughout this work. Therefore, a point of the convex set (\ref{convex}), rewritten here as
\[
\mathcal{S}= \left\{ (x,y) \in \mathbb R^2;  0 \leq \vert r_{14} \vert + \vert r_{23} \vert \leq \tfrac12 \right\},
\]
determines the matrix elements of a density operator $\rho_X$ that shares with a two-qubit state (or a family of two-qubit states) $\rho$ both the spectrum and the entanglement. Geometrically, $\mathcal S$ is nothing more than the right-triangle shown in Figure~\ref{Fig01}. Within the geometric formulation presented in this work, the populations of $\rho_X$ are assumed to be given. Therefore, the coherences $r_{14}$, $r_{23}$ and their complex conjugates determine not only the eligibility of $\rho_X$ and $\rho$ as quantum states but also the possibility of finding entanglement.

The applicability of our method goes far beyond the time-independent systems typically discussed in the literature. We have also studied what happens to entanglement when the system depends explicitly on time. Within the geometric representation, the time-evolution of $\rho_X$ describes paths  that transit between areas with different entanglement strength, even invading regions where the state becomes separable to return to the entanglement zones, and so on. The latter opens the possibility of manipulating and controlling entanglement in two-qubit systems by solving the quantum inverse problem \cite{Fer97}, where one seeks to manipulate systems in order to force them to behave in a particular way. Recent results concerning global and genuine entanglement in three-qubit systems can be found in \cite{Lun24}, for other applications see \cite{Emm00,Mie04,Cru06,Cru07,Enr17,Enr18,Lia24}.

We have found another outstanding result for the time-dependent system studied here. The envelope formed by the minima of $S$, the entropy of the reduced states, establishes an upper bound for the entanglement measures. To our knowledge, this surprising property has not been reported in the literature on the matter. 

%---------------------------------------> Section
\appendix
\section{}
\label{ApA}

\renewcommand{\thesection}{A-\arabic{section}}
% redefine the command that creates the equation no.
\setcounter{section}{0}  % reset counter 

\renewcommand{\theequation}{A-\arabic{equation}}
% redefine the command that creates the equation no.
\setcounter{equation}{0}  % reset counter 

{\bf Proposition~Q} (proof). Let us write the density operator in the form $\rho = \sum_{k,j =1}^4 r_{kj} \, \Gamma^{kj}$, where the dyadic operators $\Gamma^{kj} = \vert e_k \rangle \langle e_j \vert$, $k,j \in \{1,2,3,4\}$, correspond to 4-square matrices for which all the entries are zero except the one in the $k$th row and the $j$th column, which acquires the value 1 \cite{Enr13}. These operators obey the multiplication rule $\Gamma^{ij} \Gamma^{km} = \delta_{jk} \Gamma^{im}$, and have the properties $\sum_{k=1}^4 \Gamma^{kk} = \mathbb I$, $( \Gamma^{kj} )^{\dagger} = \Gamma^{jk}$, where $\mathbb I$ is the identity operator in $\mathcal H$. It is a matter of substitution to verify the expressions $\rho^2 = \sum_{k,j=1}^4 s_{km} \Gamma^{km}$ and $s_{km} = \sum_{\ell =1}^4 r_{k\ell} r_{\ell m}$. Then, using the factorization $r_{kj} = \alpha_k \alpha^*_j$, together with the normalization of $\vert \psi \rangle$, we obtain $\rho^2 = \rho$. Therefore $\operatorname{Tr} \rho^2 = 1$. $\blacksquare$

$\bullet$ The eigenvalues and eigenvectors of $\rho_X$ discussed in Section~\ref{Xstates} are given by the expressions
\begin{equation}
\begin{array}{c}
\Lambda_{1} = \frac12 \left[ r_{11} + r_{44} + \sqrt{ (r_{11} + r_{44})^{2} - 4 \left( x_0^2 -x^2 \right) } \right],\\[2ex]
\Lambda_{2} = \frac12 \left[ r_{11} + r_{44} -  \sqrt{ (r_{11} + r_{44})^{2} - 4 \left( x_0^2 -x^2 \right) } \right],\\[2ex]
\Lambda_{3} = \frac12  \left[ r_{22} + r_{33} + \sqrt{ (r_{22} + r_{33})^{2} - 4 \left( y_0^2 - y^2 \right) } \right],\\[2ex]
\Lambda_{4} = \frac12  \left[ r_{22} + r_{33} - \sqrt{ (r_{22} + r_{33})^{2} - 4 \left( y_0^2 - y^2 \right) } \right],
\end{array}
\label{lambdas}
\end{equation}
and
\begin{equation}
\begin{array}{c}
\vert \epsilon_{1} \rangle = \displaystyle\frac{x e^{i \theta} \vert e_1 \rangle + ( \Lambda_{1} - r_{11} ) \vert e_4 \rangle}{\sqrt{ ( \Lambda_{1} - r_{11} )^{2} + x^2 }} , \quad
\vert \epsilon_{2} \rangle = \frac{x e^{i \theta} \vert e_1 \rangle + ( \Lambda_{2} - r_{11} ) \vert e_4 \rangle }{\sqrt{ ( \Lambda_{2} - r_{11} )^{2} + x^2 }} , \\[3.5ex]
\vert \epsilon_{3} \rangle = \displaystyle\frac{y e^{i \phi} \vert e_2 \rangle + ( \Lambda_{3} - r_{22} ) \vert e_3 \rangle}{\sqrt{ ( \Lambda_{3} - r_{22} )^{2} +  y^2}} , \quad 
\vert \epsilon_{4} \rangle = \frac{ y e^{i \phi} \vert e_2 \rangle + ( \Lambda_{4} - r_{22} ) \vert e_3 \rangle}{\sqrt{ ( \Lambda_{4} - r_{22} )^{2} + y^2 }} .
\end{array}
\label{evec}
\end{equation}
In the above expressions we have used the notation introduced in Section~\ref{Xstates} for the convex set $\mathcal S$. Namely, $x_{0} = \sqrt{r_{11} r_{44}}$, $r_{14} =x e^{i \theta}$, and $y_{0} = \sqrt{r_{22} r_{33}}$, $r_{23} = y e^{i \phi}$.

$\bullet$ Let us review the situation when the matrix-elements of $\rho_X$ fulfill Proposition~Q. That is, when the X--state is pure. To satisfy (\ref{rule1}), given that $r_{13}$ and $r_{34}$ have $\alpha_3$ as a common factor, it will be enough to take $\alpha_3=0$ so that these two matrix-elements are equal to zero. Similarly, we take $\alpha_2=0$ to cancel $r_{24}$ and $r_{12}$. These $\alpha$-parameters also give $r_{22} = r_{33} =0$, so the second identity of Eq.~(\ref{rule2}) is automatically satisfied. Furthermore, we immediately obtain $r_{23}=0$. Then, the X--state (\ref{rhox}) is reduced to the projector $\rho_X = \vert \psi_{14} \rangle \langle \psi_{14} \vert$, with
\be
\vert \psi_{14} \rangle = \alpha_1 \vert e_1 \rangle + \alpha_4 \vert e_4 \rangle, \quad \vert \alpha_1 \vert^2 + \vert \alpha_4 \vert^2 =1.
\label{res1}
\ee
If we also demand that the first identity of Eq.~(\ref{rule2}) holds, then $\vert \alpha_1 \vert = \vert \alpha_4 \vert$ and $\vec \lambda_R =  (\frac12, \frac12)$. After adjusting the phases and requiring orthogonality, from (\ref{res1}) we recover the first pair of Bell--states (\ref{bell1}). Proceeding in a similar way, now eliminating $\alpha_1$ and $\alpha_4$, we obtain another projector $\rho_X = \vert \psi_{23} \rangle \langle \psi_{23} \vert$, where
\be
\vert \psi_{23} \rangle = \alpha_2 \vert e_2 \rangle + \alpha_3 \vert e_3 \rangle, \quad \vert \alpha_2 \vert^2 + \vert \alpha_3 \vert^2 =1.
\label{res2}
\ee
In this case, the second identity of Eq.~(\ref{rule2}) gives $\vec \lambda_L =  (\frac12, \frac12)$, so that (\ref{res2}) leads to the second pair of Bell--states (\ref{bell2}).

$\bullet$ The convex optimization discussed in Section~\ref{optimize} requires Lagrange multipliers. Consider the function $f(r_{11},r_{22},r_{33},r_{44}) = \sqrt{r_{11} r_{44}} + \sqrt{r_{22} r_{33}}$, together with the constraint 
\begin{equation}
g(r_{11},r_{22},r_{33},r_{44}) = r_{11} + r_{22} + r_{33} + r_{44} - 1 = 0.
\label{constr}
\end{equation}
Using $m$ as the Lagrange multiplier associated with constraint $g$, we have the system
\[
\begin{array}{c}
\frac{\partial f}{\partial r_{11}} - m \frac{\partial g}{\partial r_{11}} = \frac{r_{44}}{2\sqrt{r_{11} r_{44}}} - m = 0, \quad
\frac{\partial f}{\partial r_{22}} - m \frac{\partial g}{\partial r_{22}} = \frac{r_{33}}{2\sqrt{r_{22} r_{33}}} - m = 0,\\[2ex]
\frac{\partial f}{\partial r_{33}} - m \frac{\partial g}{\partial r_{33}} = \frac{r_{22}}{2\sqrt{r_{22} r_{33}}} - m = 0, \quad 
\frac{\partial f}{\partial r_{44}} - m \frac{\partial g}{\partial r_{44}} = \frac{r_{11}}{2\sqrt{r_{11} r_{44}}} - m = 0.
\end{array}
\]
Therefore
\[
m = \frac{r_{11}}{2\sqrt{r_{11} r_{44}}} = \frac{r_{22}}{2\sqrt{r_{22} r_{33}}} = \frac{r_{33}}{2\sqrt{r_{22} r_{33}}} = \frac{r_{44}}{2\sqrt{r_{11} r_{44}}}.
\]
The simplest way to satisfy these equalities is by making $r_{11} = r_{44}$ and $r_{22} = r_{33}$. In this case $f=r_{11} + r_{22}$. Then, from (\ref{constr}) it is found that $1/2$ is the maximum value we are looking for.

$\bullet$ The rank of states $\rho_{\mathcal M_x}$ that are represented by matrices (\ref{otros}) is as follows
\be
\left. \operatorname{rank} (\rho_{\mathcal M_x}) \right\vert_{y_0=0} = \left\{
\begin{array}{rl}
3, & r_{22}=0 \, \operatorname{or} \,  r_{33} =0\\
2, & r_{22}=0 \, \operatorname{and} \,  x=x_0\\
2, & r_{33}=0 \, \operatorname{and} \,  x=x_0\\
2, & r_{22}= r_{33} =0\\
1, & r_{22}= r_{33} =0 \, \operatorname{and} \,  x=x_0\\
\end{array}
\right.
\label{rankA}
\ee
In turn, the rank of states $\rho_{\mathcal M_y}$ that are represented by matrices (\ref{forms1}) reads
\be
\left. \operatorname{rank} (\rho_{\mathcal M_y}) \right\vert_{x_0 =0}= \left\{
\begin{array}{rl}
3, & r_{11}=0 \, \operatorname{or} \,  r_{44} =0\\
2, & r_{11}=0 \, \operatorname{and} \,  y=y_0\\
2, & r_{44}=0 \, \operatorname{and} \,  y=y_0\\
2, & r_{11}= r_{44} =0\\
1, & r_{11}= r_{44} =0 \, \operatorname{and} \,  y=y_0\\
\end{array}
\right.
\label{rankB}
\ee

%---------------------------------------> Section
\section*{Author Contributions}

Conceptualization O.R.-O.; methodology, formal analysis, investigation, original draft preparation and review, S.L.-H., C.Q. and O.R.-O.; editing, project administration and funding acquisition, O.R.-O. All authors have read and agreed to the published version of the manuscript.

%---------------------------------------> Section
\section*{Funding}

This research was funded by Consejo Nacional de Humanidades, Ciencia y Tecnolog\'ia (CONACHyT, Mexico), grant number A1-S-24569, and by Instituto Polit\'ecnico Nacional (IPN, Mexico), project SIP20242277.

%---------------------------------------> Section
\section*{Acknowledgment}

S.L.-H acknowledges the support from CONAHCyT through the scholarship 592045.

%---------------------------------------> Bibliography

\end{document}